\documentclass[twocolumn,showpacs,preprintnumbers,amsmath,amssymb,endfloats*]{revtex4}
\usepackage{graphicx}

\begin{document}

\thispagestyle{empty}

\title{Comment on ``Temperature dependence of the Casimir force
for lossy bulk media''}

\author{G.~Bimonte,${}^1$
G.~L.~Klimchitskaya,${}^{2,3}$ and
V.~M.~Mostepanenko${}^{3,4}$}

\affiliation{${}^1$Dipartimento
di Scienze Fisiche, Universit\`{a}
di Napoli Federico II, Complesso
Universitario MSA, Via Cintia
I---80126 Napoli, Italy \\
INFN, Sezione di
Napoli, Napoli, Italy \\
${}^2${North-West Technical University,
Millionnaya St. 5, St.Petersburg,
191065, Russia}\\
${}^3$Department of Physics, Federal University of Para\'{\i}ba,
C.P.5008, CEP 58059--900, Jo\~{a}o Pessoa, Pb-Brazil \\
${}^4${Noncommercial Partnership ``Scientific Instruments'',
Tverskaya St. 11, Moscow,
103905, Russia}
}

\begin{abstract}
Recently Yampol'skii {\it et al.} [Phys. Rev. A {\bf 82}, 032511
(2010)] advocated that Lifshitz theory is not applicable when the
characteristic wavelength of the fluctuating electromagnetic
field, responsible for the thermal correction to the Casimir
force, is larger than the size of the metal test bodies. It was
claimed that this is the case in experiments which exclude
Lifshitz theory combined with the Drude model. We calculate the
wavelengths of the evanescent waves giving the dominant
contribution to the thermal correction and we find that they are
much smaller than the sizes of the test bodies. The opposite
conclusion obtained by the authors arose from a confusion between
propagating and evanescent waves.
\end{abstract}
\maketitle

It is the subject of a considerable body of literature that
theoretical predictions for the thermal
Casimir force between lossy metal plates described by the Drude
model, based on Lifshitz theory,
are in disagreement with experimental data (see, e.g.,
review \cite{1}). In Ref.~\cite{2} an attempt is undertaken to
explain this contradiction by arguing that Lifshitz theory is
indeed inapplicable to  test bodies of finite size, such as those
used in the experiments. According to Ref.~\cite{2},  the thermal
electromagnetic fluctuations responsible for the predicted large
thermal correction \cite{3},  excluded by several experiments,
have a characteristic wavelength which is larger than the size of
the test bodies used in the experiments. On this basis a
conclusion is made that the predicted correction can be observed
experimentally only for sufficiently large metal bodies. Below we
show that the wavelengths of the fluctuations contributing to the
large thermal correction  engendered by the Drude model, are in
fact much less than the sizes of test bodies used in related
experiments. Because of this, the purported explanation of the
contradiction between experiment and theory in Ref.~\cite{2} is in
error. We argue that the considerable overestimate made in
Ref.~\cite{2} of the  wavelengths of the contributing
fluctuations, was the result of a confusion between travelling
(propagating) and evanescent waves.

The frequencies and wave-vectors of the fluctuating
electromagnetic field giving  a major contribution to the thermal
correction to the Casimir force can be found using Lifshitz
formula written in terms of real frequencies. In modern notation,
the thermal correction to the Casimir force per unit area, between
two parallel semispaces at temperature $T$ separated by a gap of
width $l$, can be represented in the form \cite{4,4a}
\begin{eqnarray}
&&
F_{\rm rad}(l)=-\frac{\hbar}{\pi^2}
\int_{0}^{\infty}k_{\bot}dk_{\bot}\int_{0}^{\infty}
\frac{d\omega}{e^{\hbar\omega/k_BT}-1}
\label{eq1} \\
&&~~~
\times
{\rm Im}\left\{q\sum_{s={\rm TM,TE}}
\left[r_s^{-2}(\omega,k_{\bot})e^{2lq}-1\right]^{-1}\right\}.
\nonumber
\end{eqnarray}
\noindent Here, $k_{\bot}=|\mbox{\boldmath$k$}_{\bot}|$ is the
magnitude of the projection of the wave vector onto the boundary
planes, $\omega$ is the wave frequency, $k_B$ is the Boltzmann
constant, and $q^2\equiv
q^2(\omega,k_{\bot})=k_{\bot}^2-\omega^2/c^2$. The reflection
coefficients for two independent polarizations of the
electromagnetic field (transverse magnetic, $s={\rm TM}$, and
transverse electric, $s={\rm TE}$) are given by
\begin{equation}
r_{\rm TM}(\omega,k_{\bot})=
\frac{\varepsilon(\omega)\,q-k}{\varepsilon(\omega)\,q+k},\quad
r_{\rm TE}(\omega,k_{\bot)}= \frac{q-k}{q+k}, \label{eq2}
\end{equation}
\noindent
where
\begin{equation}
k^2\equiv k^2(\omega,k_{\bot})=k_{\bot}^2-\varepsilon(\omega)
\frac{\omega^2}{c^2},
\label{eq3}
\end{equation}
\noindent and $\varepsilon(\omega)$ is the frequency-dependent
dielectric permittivity of the material of the semispaces.
Equation (\ref{eq1}) coincides with Eq.~(3) of Ref.~\cite{2},
after correcting one misprint contained there  (in the exponent in
the Boltzmann factor on the right-hand side of Eq.~(3) the factor
of two should be erased; an analogous misprint should be corrected
in Eq.~(5) of Ref.~\cite{2}).

In Ref.~\cite{5} it was shown that if the material of the
semispaces is described by the Drude model
\begin{equation}
\varepsilon(\omega)=1-\frac{\omega_p^2}{\omega(\omega+i\nu)},
\label{eq4}
\end{equation}
\noindent
where $\omega_p$ is the plasma frequency and $\nu$ is the
relaxation parameter, the major contribution to $F_{\rm rad}$
 is given by TE {\it evanescent} waves.
For example, for Au semispaces with $\hbar\omega_p=9\,$eV and
$\nu=5.32\times10^{13}\,$rad/s at a separation  $l=162\,$nm, the
TE evanescent waves contribute  about 99.7\% of the thermal
correction. This contribution can be  denoted as $F_{\rm
rad,\,TE}^{\rm evan}$. In Eq.~(\ref{eq1}), the quantity $F_{\rm
rad,\,TE}^{\rm evan}$ is obtained by taking the term with $s={\rm
TE}$, for frequencies $\omega$ varying in the interval from 0 to
$ck_{\bot}$, for which the quantity $q$ is real.  Even though the
concept of evanescent waves is never mentioned in Ref.~\cite{2},
the contribution of TE evanescent waves is actually reproduced by
the quantity in the first pair of square brackets in Eq.~(3) of
Ref.~\cite{2}, integrated over imaginary values of $p$ ranging
from $i0$ to $i\infty$. The same contribution can be physically
interpreted in terms of interaction of eddy currents \cite{5a,5b}.

As it was also shown in Ref.~\cite{5}, at short separations
between two  semispaces  described by Eq.~(\ref{eq4}), the
frequencies $\omega$ giving a dominant contribution to $F_{\rm
rad,\,TE}^{\rm evan}$ satisfy the inequality
$\omega\lesssim\nu(\omega_c/\omega_p)^2$ where $\omega_c=c/(2l)$
is the characteristic frequency. This result was qualitatively
confirmed in Ref.~\cite{2} where the frequencies contributing to
the quantity $F_{\rm rad,\,TE}^{\rm evan}-\left. F_{\rm
rad,\,TE}^{\rm evan}\right|_{\nu=0}$ were found to satisfy the
inequality $\omega\lesssim\nu$ (at $l=100\,$nm, it holds
$\omega_c\approx\omega_p/9$). We note that the term $\left. F_{\rm
rad,\,TE}^{\rm evan}\right|_{\nu=0}$, which is subtracted from
$F_{\rm rad,\,TE}^{\rm evan}$ in Ref.~\cite{2}, represents a
negligibly small thermal effect that results once the material for
the semispaces is described by the plasma model,  and it does not
influence any of the obtained conclusions.

It is important to realize that the characteristic wavelength of
the evanescent waves giving the largest contribution to the
thermal correction is determined, however, not by the frequency
spectrum of $F_{\rm rad,\,TE}^{\rm evan}$, but rather by its {\it
wave-vector spectrum}. In order to determine the latter spectrum
numerically, we have recast the quantity $F_{\rm rad,\,TE}^{\rm
evan}$ in the following equivalent form in terms of dimensionless
variables
\begin{equation}
F_{\rm rad,\,TE}^{\rm evan}(l)=\frac{\hbar\,\nu\,
c^2}{\pi^2\omega_p^2\,l^5} \int_{0}^{\infty} dvv^2g(v),
\label{eq5}
\end{equation}
\noindent
where
\begin{equation}
g(v)=\int_{0}^{\infty}
\frac{du}{\exp\left(\frac{\hbar\nu}{k_BT}\frac{c^2}{\omega_p^2l^2}u\right)-1}
\,{\rm Im}\!\left[1-\frac{e^{2v}}{r_{\rm TE}^2(u,v)}\right]^{-1}.
\label{eq6}
\end{equation}
\noindent
Here, the new variables are defined as
\begin{equation}
u=\frac{\omega_p^2l^2}{\nu c^2}\omega, \quad
v=lq.
\label{eq7}
\end{equation}
\noindent
In terms of these variables the TE reflection coefficient is
given by
\begin{equation}
r_{\rm TE}(u,v)=\frac{v-\sqrt{v^2+
\frac{\omega_p^2l^2u}{i\omega_p^2l^2+c^2u}}}{v+\sqrt{v^2+
\frac{\omega_p^2l^2u}{i\omega_p^2l^2+c^2u}}}.
\label{eq8}
\end{equation}

We have computed the range $v_1\leq v\leq v_2$ of the variable $v$
which contributes 90\% of $F_{\rm rad,\,TE}^{\rm evan}$ at the
experimental separation $l=162\,$nm. For $\nu=5.32\times
10^{13}\,$rad/s, we found $v_1=0.26$ and $v_2=3$, while for
$\nu=10^{10}\,$rad/s we obtained $v_1=0.28$ and $v_2=3$. Thus,
independently of the values of the relaxation parameter $\nu$, the
dimensionless quantity $v$ contributing to $F_{\rm rad,\,TE}^{\rm
evan}$ is always of order one, and therefore $q$ is always of
order $1/l$.

The wave-vector   of an evanescent wave is given by the expression
\begin{equation}
\mbox{\boldmath$k$}=(k_x,k_y,k_z), \quad
k_z=\sqrt{\frac{\omega^2}{c^2}-k_{\bot}^2}=iq,
\label{eq9}
\end{equation}
\noindent and its wavelength is determined as
\begin{equation}
\lambda=\frac{2\pi}{k_{\bot}}=\frac{2\pi}{\sqrt{k_x^2+k_y^2}}.
\label{eq10}
\end{equation}
\noindent Keeping in mind the definition of $q$, we then obtain
that for the most contributing wave-vectors it holds
$k_{\bot}^2>q^2\sim 1/l^2$, in such a way that the corresponding
wavelengths satisfy the inequality
\begin{equation}
\lambda\lesssim 2\pi \,l\;.
\label{our}
\end{equation}
\noindent
In the experiments aiming at measuring the
Casimir force between a sphere and a plate, these wavelengths are
always much smaller than the characteristic size $L$ of the part
of the sphere surface
\begin{equation}L\approx 2\sqrt{R^2-(R-l)^2}\approx 2\sqrt{2Rl},
\label{eq10a}
\end{equation}
\noindent which contributes to the force. For example, in the
experiment of Ref.~\cite{6} the sphere radius is $R=150\,\mu$m,
and the separation distances vary from $l=162\,$nm to
$l=750\,$nm.  For such values of $R$ and $l$, the inequality
$\lambda_{\max}=2\pi l<L$, i.e., $l<2R/\pi^2\approx 30\,\mu$m is
satisfied with large safety margins, for all the separations
considered. In fact, the relevant contributing wavelengths are
smaller than the sizes of the bodies, in all other experiments
measuring the Casimir force performed up to date as well \cite{1}.

The opposite conclusion  obtained in Ref.~\cite{2} is caused by a
confusion between  propagating and evanescent waves. {} Starting
from a qualitatively correct inequality for the contributing
frequencies $\omega\lesssim\nu$, the authors of Ref.~\cite{2}
  used the  following relation between the frequency and the
period
\begin{equation}
\omega=\frac{2\pi c}{\lambda},
\label{eq11}
\end{equation}
\noindent  to obtain the estimate  $\lambda \gtrsim 2\pi c/\nu$
for the wavelengths of the fluctuations contributing to $F_{\rm
rad,\,TE}^{\rm evan}$. Thereafter, it was concluded that Lifshitz
theory is only applicable if the size of test bodies $L\gg 2\pi
c/\nu$, i.e., $\nu\gg 2\pi c/L$ (Eq.~(9) in Ref.~\cite{2}). The
problem with this  argument, though,  is that Eq.~(\ref{eq11})
is valid only for travelling (propagating) waves in
vacuum. In this case the two definitions of the wavelength
$\lambda=2\pi c/\omega=2\pi/|\mbox{\boldmath$k$}|$ coincide.
Unfortunately, in the case of evanescent waves, which do not
propagate and are more similar to standing waves, Eq.~(\ref{eq11})
does not hold, and the wavelength has no relation to the
frequency. If instead of using Eq.~(\ref{eq11}), the authors of
Ref.~\cite{2} had considered the characteristic values of their
parameter $x=2i\,p\,\omega \,l/c$ (where in our notation
$p=-iqc/\omega$) to determine the most contributing wavelengths,
our result $\lambda\lesssim 2\pi l$ would have been reproduced.
Indeed, as shown in Ref.~\cite{2}, $x=2\,l\,q\sim 1$ leading to
$q\sim 1/(2\,l)$ in qualitative agreement with our estimate
(\ref{our}).  Bearing in mind that for evanescent waves the
frequency is unrelated to the wavelength, the second inequality,
$\omega\lesssim k_BT/\hbar$, considered in Ref.~\cite{2} does not
lead to any constraint on the size of bodies $L$. For the same
reason the results of numerical computations presented in Figs.~1
and 2 of Ref.~\cite{2} do not contain any information concerning
the role of finite sizes of the test bodies in calculations of the
thermal Casimir forces.

To conclude, the problem of the disagreement between the
experimental data of several experiments and the theoretical
prediction of the thermal effect in the Casimir force, obtained by
using Lifshitz theory  in combination with the Drude model,
remains unsolved.

\section*{Acknowledgments}

G.B.~thanks ESF Research Network CASIMIR for financial support.
G.L.K.\ and V.M.M.\ are grateful to the Federal University of
Para\'{\i}ba for kind hospitality.  They were supported
by CNPq (Brazil).

\end{document}